\documentclass[iop]{emulateapj}

\usepackage{epsfig} 
\usepackage{amsmath}
\usepackage{amssymb}
\usepackage{natbib}
\usepackage{graphicx}
\usepackage{gensymb}
\usepackage{lineno}
\usepackage{endnotes}

\newcommand{\nustar}{{\it NuSTAR}}
\newcommand{\xmm}{{\it XMM-Newton}}
\newcommand{\suzaku}{{\it Suzaku}}
\newcommand{\sax}{{\it BeppoSAX}}
\newcommand{\xte}{{\it RXTE}}
\newcommand{\swift}{{\it SWIFT}}
\newcommand{\maxi}{{\it MAXI}}
\newcommand{\chandra}{{\it Chandra}}

\shorttitle{anamolous soft spectral state of V4641 Sgr}
\shortauthors{Pahari et al.}

\begin{document}

\title{Constraining distance and inclination angle of V4641 Sgr using {\it SWIFT}  and {\it NuSTAR} observations during low soft spectral state}
\author{Mayukh Pahari$^1$, Ranjeev Misra$^1$, Gulab C Dewangan$^1$, Pramod Pawar$^{1,2}$ 
}
\affil{$^1$ Inter-University Centre for Astronomy and Astrophysics, Pune, India; \texttt{mayukh@iucaa.ernet.in} \\}
\affil{$^2$ Swami Ramanand Teerth Marathwada University, Nanded, India \\}

\begin{abstract}

We present results from \nustar{} and \swift{}/XRT joint spectral
analysis of V4641 Sgr during a disk 
dominated or soft state as well as a powerlaw dominated 
or hard state. The soft state spectrum is well modeled by a 
relativistically blurred disk emission, a powerlaw, a broad Iron
line, two narrow emission lines and two edges. The Markov Chain Monte Carlo simulation technique and the relativistic effects
seen in the disk and broad Iron line allow us to self-consistently constrain 
the inner disk radius, disk
inclination angle and distance to the source at 2.43$^{+0.39}_{-0.17}$
R$_g$ (GM/c$^2$), 69.5$^{+12.8}_{-4.2}$ degrees and 10.8$^{+1.6}_{-2.5}$ kpc respectively.
For the hard state, the spectrum is a power-law with a weakly broad
Iron line and an edge. The distance estimate gives a measure of the Eddington fraction,
$L_{2.0 - 80.0 keV}/L_{Edd}$, to be $\sim$1.3 $\times$ 10$^{-2}$ and $\sim$1.9 $\times$ 10$^{-3}$ for the
soft and hard states respectively. Unlike many other typical black hole
systems which are always in a hard state at such low Eddington fraction,
V4641 Sgr shows a soft, disk dominated state.
The soft state spectrum shows narrow emission lines at $\sim 6.95$ and $\sim 8.31$ keV which can
be identified as being due to emission from highly ionized  Iron and Nickel 
in an X$-$ray irradiated wind respectively. If not due to instrumental effect
or calibration error, this would be the first detection of
a Ni fluorescent line in a black hole X$-$ray binary.

\end{abstract}

\keywords{accretion, accretion disks --- black hole physics ---
X-rays: binaries --- X-rays: individual (V4641 Sgr)}

\section{Introduction}

Relativistically skewed Fe line complex at $\sim$6.5 keV is a
frequently observed phenomenon in X$-$ray spectra of a range of objects $-$ from
Seyfert galaxies \citep{fa00} to low mass X$-$ray binaries harboring neutron stars and black holes \citep{ca08,ko14}. These features
reveal the General Relativistic nature of these objects and 
can be used to probe the space-time metric near black holes.
Such broad lines have also been reported for Galactic X$-$ray binaries
which harbour either a black hole or a Neutron star. 
Among these Galactic black hole X$-$ray binaries (BHXBs), till
now, few transient sources - GRO J1655-40 \citep{to99}, XTE J1550-564
\citep{so00}, XTE J1650-500 \citep{do06}, XTE J1752-223 \citep{re11},
SWIFT J1753.5-0127 \citep{hi09}, SAX J1711.6-3808 \citep{in02}, GX
339-4 \citep{fe01} and two persistent sources - Cyg X-1
\citep{mi02b,du11} and GRS 1915+105 \citep{mi13}, have shown strong,
relativistically broad Fe K$\alpha$ lines.
At lower luminosities, broad Iron line has
been reported only for two sources GX 339-4
\citep{mi06} and XTE J1650-500 \citep{do06}.

BHXBs show a variety of spectral states, each with
its distinct timing properties \citep[e.g.][]{re06}. These states
have been primarily classified based on the outburst behavior of 
transient black hole systems with low mass companions i.e. low mass black
hole X$-$ray binaries (LMXBs). These outbursts which typically last for 
few months, often trace out a `q' shaped figure in the luminosity hardness
diagram \citep{ho01,fe04}. 
The outburst starts at right hand bottom part of the diagram,
where the source has low luminosity (L/L$_{Edd}$ $<$ 0.01) and is hard. Then roughly maintaining
its hardness ratio, the source increases its luminosity till it reaches
near the peak of the outburst (L/L$_{Edd}$ $>$ 0.3). From this point, at nearly similar luminosity level,
the sources softens, i.e., it  moves horizontally left in the luminosity hardness
diagram. From this Eddington limited soft state, the source 
luminosity decreases as the outburst decays while maintaining its soft nature,
 till the luminosity falls
to roughly $0.01$-$0.1$ times of its Eddington value. From this point onward the
source makes a transition from soft to hard state i.e. it moves right
in the luminosity hardness diagram. After reaching the hard state the
source  luminosity starts dropping further till it reaches 
its quiescent value (L/L$_{Edd}$ $\sim$ 10$^{-5}-$10$^{-3}$). Broadly
speaking, in the soft state (especially in the low luminosity soft states)
the spectrum is dominated by a standard accretion disk extending to the
innermost stable orbit close to the black hole. While in the hard
state (especially in the low luminosity hard state) the disk emission
is weak probably because it is truncated at a large radius away
from the black hole and the spectrum is dominated by the hard power-law
from an hot inner corona. Now, the origin of relativistically broadened Iron line is usually 
associated with a standard accretion disk extending close to the black hole. hence its surprising that such lines have
been detected during the hard states of GX 339-4 and XTE J1650-500. However, in an alternate scenario, it has been observed that the cold material rotating close to central black hole (e.g., Cyg X$-$1; \citet{fa89}) can be illuminated by the primary power-law photons to give rise the broad Iron emission line \citep{ba01}.

While the transition from soft to hard occurs at different luminosity levels,
no source has ever shown a soft state for luminosities below $< 0.01 L_{Edd}$.
At these luminosities, all black hole binaries show hard spectrum with
very weak disk emission. 

Black Hole systems with high mass companions i.e. high mass
X$-$ray binaries (HMXBs) are typically persistent sources with 
luminosities ranging from $0.01$ to $0.1 L_{Edd}$. They also
show few  spectral states which can be broadly classified into hard and soft
but not as many as those of the LMXBs. This is expected since a persistent
HMXB  varies over a much narrower range of luminosity. Like LMXBs, in 
the hard state, they show a week disk emission while in the soft state
the disk emission is significantly stronger although an extreme soft state,
where the disk totally dominates the emission is rarely seen in HMXBs.
Hence, HMXBs should also show broad Iron lines in the soft state
and this is seen, e.g., in Cyg X-1 and also in LMC X-1. A fundamental difference 
between the LMXBs and HMXBs is the presence of a strong wind
from the companion and one of the consequences of this wind is
the X$-$ray irradiation on it can produce emission lines and edges which
can appear in the Iron line band. These features carry important 
information regarding the structure and composition of the wind, but
when observed through low spectral resolution instruments, they may
lead to improper modeling of the broad Iron line originating from the
disk. 

Discovered during the historical outburst with an intensity of 12.2
Crab by \xte{}/ASM, V4641 Sgr reveals itself as the only transient
high mass black hole binary (HMBHXB) with a dynamically confirmed
black hole mass of 6.4 $\pm$ 0.4 M$\odot$ and a companion mass of
2.9 $\pm$ 0.4 M$\odot$ \citep{ma14}. Using ellipsoidal model
fitting to the photometric data during an optical passive state,
\citet{ma14} determined a disk inclination angle to be  ({\it i}) of 72.3
$\pm$ 4 $\degree$. Prior to this, a qualitative argument favoring
{\it i} $\le$ 10$\degree$, was provided in order to explain the high beamed
jet expanding at a velocity between 0.22 and 1 $\arcsec$ per day
\citep{hj00}. In another study, based on high amplitude folded light
curve analysis, the inner disk inclination angle was derived between
60$\degree$ and 70$\degree$ and the distance to the source was predicted
to lie between 7 kpc and 12 kpc \citep{or01}. 
Fitting the \chandra{} spectra from this source in the energy range
0.5$-$10.0 keV during the quiescent state requires partial covering
absorption and distant reflection \citep{mo14}, similar to model
components require to fit spectra from Seyfert$-$2 AGNs. A strong Fe
emission line from this source was first detected by \sax{}
\citep{in00} with the equivalent width between 0.3 and 1 keV. Later,
using \xte{} data, the presence of a Compton reflection hump around
20$-$30 keV along with strong Fe line have been observed
\citep{ma06}. 

\citet{mi02a}  re-analyzed the \sax{}/MECS data, and reported
a broad Fe K$\alpha$ emission line complex  and preferred it over a model 
consisting of two narrow emission lines. Given that one may expect narrow
emission lines from the wind, it is important to measure the spectra
of V4641 Sgr with a high resolution instrument. Moreover, the uncertainties
in its distance and inclination angle do not allow it to be compared
to other black hole systems. An independent distance analysis is required to
understand whether this HMXB, at a certain Eddington luminosity ratio,
has the same or different spectral shape compared to others.

Till the advent of \nustar{} satellite, \xmm{}, \chandra{} and
\suzaku{} were the only instruments for sensitively 
 measuring broad relativistic Fe
line although  they were highly restricted by their pile-up limit. 
With its
pile-up free operation (up to $\sim$100 mCrab), good calibration 
and high energy resolution ($\sim$ 400 eV at 0.1$-$10 keV), \nustar{} provides a unique
opportunity to study relativistically skewed fluorescent line profile accurately and to constrain other features of the reflection component in the 3.0$-$79.0 keV energy band \citep{ha13}.  For example, a 15 ks
exposure of GRS 1915+105 reveals an extremely skewed Fe emission line
which allowed for an estimation of the black hole spin to be 0.98 $\pm$ 0.01 \citep{mi13}. \nustar{} has already shown promising results regarding detection of lines other than Fe K$\alpha$. For example,  Fe K$\beta$ line is observed from the Compton thick Active Galactic Nuclei (AGN) $-$ NGC 424, NGC 1320 and IC 2560 using both \xmm{} and \nustar{} spectra \citep{ba14}. The truly complex nature of the X$-$ray spectra in BHXBs are often demonstrated by high resolution spectroscopic instruments and attentions are required while modeling if truly robust information are needed to obtain. For example, reanalysis of the complex \suzaku{} spectra in the Seyfert 1.5 galaxy NGC 3783 using Markov Chain Monte Carlo (MCMC) simulation technique shows that the data strongly requires both super$-$solar Iron abundance as well as rapidly spinning black hole \citep{re12} which was inconclusive from previous attempts that uses conventional methods. Therefore, although the good spectral resolution of \nustar{} provides a unique opportunity to discern narrow emission lines and edges while simultaneously measuring the broad Iron line profile, proper statistical analysis like MCMC simulations of the best$-$fit model is needed for the bias-free determination of probable inter-dependent parameters from different model components.  

\nustar{} has observed  V4641 Sgr twice, once when the
source was soft and the other when it was hard. We analyze
these observations along with quasi-simultaneously \swift{} data.
Our motivation is to use the broad band $0.5$-$70$ keV data to
constrain the disk emission, the hard X$-$ray powerlaw, the
 broad Iron line as well as any narrow emission features that may arise from the wind. As we shall
see, simultaneous measurement of the disk emission and the broad Iron
line will allow us to constrain the distance to the source as
well as its inclination angle. Observation details and data reduction procedure are provided in section 2. Spectral analysis,
modeling and results are discussed in section 3. Section 4 provides
a comparison with other transient BHXBs. Discussion and conclusions are provided in section 5.

\section{Observations \& Data reduction} 

\begin{figure*}
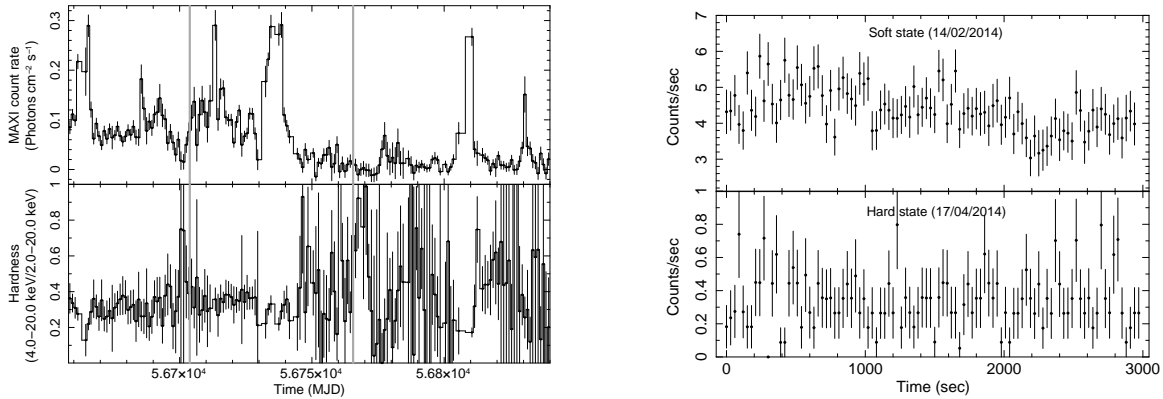

\hspace*{\fill}%
\includegraphics[scale=0.30,angle=-90,keepaspectratio]{fig1a.ps}\vspace{10mm}
\hspace*{\fill}%
\includegraphics[scale=0.29,angle=-90,keepaspectratio]{fig1b.ps}
\hspace*{\fill}%
\caption{Left panel: The 2.0$-$20.0 keV light curve and hardness of V4641 Sgr is shown as a function of
time obtained from \maxi{}/GSC. The two gray, vertical lines
mark the times of the  two simultaneous \nustar{} and \swift{}
observations analyzed in this work. Note that the state was in a
soft state in the first observation and a hard one during the second. Right panel: \nustar{}/FPMA and FPMB averaged, background-subtracted lightcurve are shown during soft state and hard state with 30 sec bintime.}
\label{maxilc}
\end{figure*}

\nustar{} observed  V4641 Sgr on 14 February,
2014 00:36:07 (80002012002) and on 17 April, 2014 22:46:07 (80002012004)
with exposure times of 24 ks and 26 ks respectively. During this period,
the source was variable as seen in the left panel of
Figure \ref{maxilc} where the $2$-$20$ keV \maxi{}/GSC light curve is shown.
The two vertical, gray lines mark the time of the \nustar{} observations
and in between one can see flaring activities. 
From \maxi{} lightcurve, a drop in count rate by a factor of $\sim$3 is observed at the time of second \nustar{} observation compared to the first. Not only that, hardness ratio, as observed from the left panel of Figure \ref{maxilc}, increases by a factor of $\sim$2 at the time of second \nustar{} observation. Background subtracted, \nustar{}/FPMA and FPMB averaged, 3 ks lightcurves with 30 sec bintime is shown in the right panel of Figure \ref{maxilc} during first and second observations respectively. The average source and background count rate from \nustar{} data during the first observation is 5.35 $\pm$ 0.02 counts/sec and 0.81 $\pm$ 0.03 counts/sec respectively while the same during second observation are 0.31 $\pm$ 0.05 and 0.042 $\pm$ 0.008 counts/sec respectively. The 0.01$-$5.0 Hz integrated rms power from the white-noise subtracted power density spectra (PDS) of the second \nustar{} observation is $\sim$3 times than that observed during first \nustar{} observation. Above facts indicate that during the first \nustar{} observation the source was in soft spectral state while during the second observation it was in hard spectral state. Later, in this work we confirm soft-to-hard spectral state transition from (1) spectral analysis where the soft disk blackbody flux dominates the spectra ($>$ 80\%) during the first \nustar{} observation while powerlaw flux dominates the spectra ($>$ 80\%) during second \nustar{} observation. (2) In the Hardness Intensity Diagram (HID), first \nustar{} observation occupy the place where other cannonical BHXBs usually show soft state while the second observation reside near the hard state position of other BHXBs.   

\nustar{} data from the co$-$aligned telescope with two focal plane
modules, FPMA and FPMB are extracted and analyzed using {\tt
nustardas v. 1.3.1}, {\tt FTools v 6.15.1} and {\tt XSPEC v.
12.8.1g}. A circular region of 40$\arcsec$ radius is used in the
detector plane to extract source spectrum and light curve. A large area
from the remaining part of the detector, which is at least thrice the size of source extraction area, is used to extract background
spectrum and light curve. \nustar{} data reduction procedure is guided
by the {\it NuSTAR Data Analysis Software Guide; v 1.7.1} provided by
{\it HEASARC} and the latest Calibration database ({\tt CALDB} version {\tt 20150702}) is used. Observation
specific response matrices and ancillary response files are generated
using {\tt numkrmf} and {\tt numkarf} tools.

\swift{}/XRT observations were also performed with 1.9 ks and 3.9 ks exposures simultaneously
with \nustar{}.
Following standard procedure of filtering and screening criteria,
\swift{}/XRT data are reduced using {\tt xrtpipeline v. 0.13.0}
tool. 30$\arcsec$ circular region is used to extract source
spectrum and 30$\arcsec$ source$-$free region is used to extract
background spectra using {\tt XSELECT v 2.4}. Latest \swift{}/XRT
spectral redistribution matrices are used. {\tt xrtmkarf} task is used
along with exposure map file to generate auxiliary response file for
the current observation.

In \nustar{} observations, background subtracted total counts during soft state and hard state are 106980 $\pm$ 327 and 5853 $\pm$ 68 respectively. We also study signal-to-noise ratio (SNR) as a function of channels. Around channel number 100, the SNR profile peaks at $\sim$5$-$6 and it falls below 1 at channel numbers less than 10 (corresponds to the energy $<$ 3 keV) and at channel numbers above 230 (corresponds to the energy $>$ 11.0 keV). Therefore, to obtain good spectral quality, data was grouped such that a minimum of 20 counts were obtained per bin. However, it is important to note that binning is not effectively applied to the largest part of the spectrum (3.0$-$15.0 keV) since background-subtracted source counts per channel $>$ 200. There were too few counts in the \nustar{} data above 30 keV and hence we restricted the analysis to 30 keV. In case of \swift{}/XRT, Background subtracted total counts during soft and hard state are 22916 $\pm$ 134 and 1855 $\pm$ 33 respectively. Because of poor signal strength in the \swift{}/XRT spectra above 6 keV (less than 10 counts), the energy range for simultaneous spectra fitting using {\tt XSPEC v 12.8.1} was restricted to 0.3$-$6.0 keV (\swift{}/XRT) and 3.0$-$30.0 keV (\nustar{}).  

\section{Spectral analysis and results}

\subsection{soft state spectral analysis}

To understand the complexities of the soft state spectra, we used a powerlaw modified by the Galactic
absorption. Throughout our analysis we obtained a Galactic absorption
column density between 0.2$-$0.4 $\times$ 10$^{22}$ cm$^{-2}$ 
which is consistent with previous results
\citep{mi02a,di90}. As it turns out to be that except powerlaw component, two basic components are required : an optically thick, soft blackbody emission which is revealed by a large excess in low energies and strong reflection features, consists of emission lines and edges between 5.8$-$10.0 keV. 

\begin{figure}
\includegraphics[scale=0.30,angle=-90,keepaspectratio]{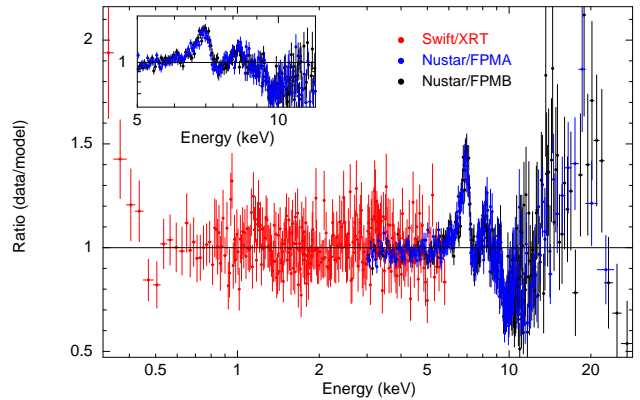}
\caption{Data to model ratio plot in the 0.3$-$30.0 keV energy range for the 
soft state spectrum  (on 14 February, 2014) of V4641 Sgr is shown using simultaneous 
\nustar{}/FPMA (black) and \nustar{}/FPMB (blue) and \swift{}/XRT (red) data.
The continuum model that has been used to calculate the ratio is a disk emission and power-law modified
by the Galactic absorption. In the inset, the Iron emission line region is shown for clarity. The residuals show a broad Iron line, a
emission line at $\sim 8$ keV and a edge at $\sim 9.5$ keV.}
\label{residual}
\end{figure}

The large residuals in the high energy region are examined in detail.
To highlight the features we show the residuals in the  5$-$11 keV band
in Figure \ref{residual}. Visual inspection of
the figure suggests several features: (i) a broad Fe line peaking
at $\sim$6.9 keV and extending perhaps to $\sim$5.5 keV, (ii) an emission
line-like feature  at $\sim$ 8.1 keV and (iii) an edge at
$\sim$9.6 keV. With these suggestive features as guidelines
we undertook a systematic analysis of the broad band data.
As compared to a base model of a disk emission and a powerlaw
component, a broad Iron line is significantly required resulting
in a $\Delta \chi^2 > 700$. We chose the relativistic line model
{\tt laor} \citep{la91} available in {\tt XSpec} to represent the relativistically blurred Iron line emission. This model
is parametrized by the line rest frame energy, an emissivity index,
inner radii in terms of $r_g = GM/c^2$, inclination angle and
normalization. We fixed the outer radius of the emission to a large
value of $400 r_g$. Like any other complex reflection models, for example, {\it reflionx} \citep{ro05} and {\it relxill} \citep{ga13} which are frequently used to describe complex reflection features (e.g., reflection hump at 20$-$30 keV) from the AGN spectra in a self$-$consistent manner, {\tt laor} model provides equally good description of the broadening of emission lines due to general relativistic effects. Since we do not have sufficient signal-to-noise ratio at energies $>$ 15 keV, reflection humps are not detected in the spectra. Therefore, spectral quality do not meet the requirement of applying relativistic reflection models more complex than {\tt laor} model. 

For consistency, we modeled the disk emission
using the relativistic disk model {\tt kerrd} \citep{eb03} in {\tt XSpec}. This disk
model uses the same formalism as the {\tt laor} model to
take into account relativistic effects. The model has convenient
parameters such as the black hole mass, which we fixed to the measured value
of  6.4 M$\odot$ and a color correction factor which we fixed
to the standard value of $1.7$ \citep{sh95}. The
other free parameters are the distance to the source and the 
accretion rate, inner disk radius and disk inclination angle. {\tt kerrd} model assumes a
Kerr metric around a spinning black hole with spin parameter
$a=0.998$ for which the last stable orbit should be at $\sim 1.235 r_g$.
A larger inner radii may signify a truncated disk or that the spin
parameter is less than what was assumed in the model, a point which
we discuss in the last section.

\begin{figure}
\includegraphics[scale=0.32,angle=-90,keepaspectratio]{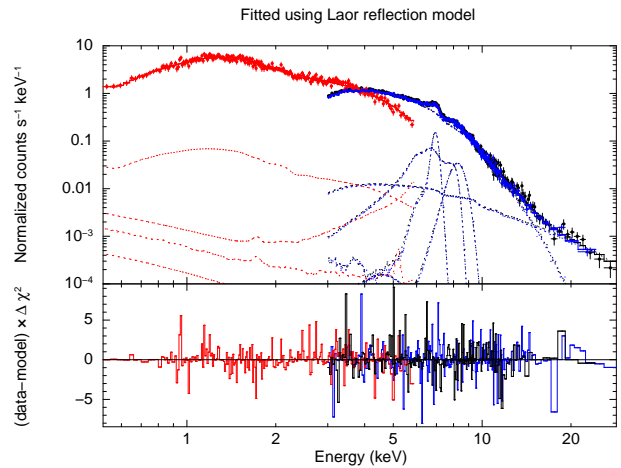} 
\caption{Simultaneously fitted \swift{}/XRT and
\nustar{}/FPMA \& FPMB energy spectra during soft state along with model
components and residual contribution to $\chi^2$ is shown. The model consists of
relativistic disk emission, powerlaw, Laor line profile, Gaussians at $\sim$6.95 keV and $\sim$8.1 keV and absorption edges at $\sim$7.1 keV and $\sim$9.6 keV.}
\label{softspec}
\end{figure}

The model consisting of a disk, powerlaw and broad line emissions
is not an adequate representation of the high quality data with
$\chi^2/dof = 1332/705$. Corresponding to broad K$_\alpha$, represented by {\tt laor} model, a K$_\beta$ line is expected with the same shape parameter. The fit significantly improved by the
addition of a narrow emission line at $\sim 6.95$ keV 
($\chi^2/dof = 807/703$) and another at $\sim 8.13$ keV 
($\chi^2/dof = 776/700$). Examination of  the residuals showed 
the presence of an absorption edge at $\sim 9.6$ keV 
($\chi^2/dof = 738/698$) and another one at  $\sim 7.1$ keV
($\chi^2/dof = 721/696$). In the fitting process previous model components were free to vary while the new one is being fitted. The best fit parameters for this model are listed in Table \ref{specpar} and the unfolded spectrum with
residuals are shown in Figure \ref{softspec}. To check the goodness of the fit, we use Kolmogorov-Smirnov test statistic that uses Empirical Distribution Function (EDF) and perform goodness of fit Monte Carlo simulation to check what percentage of these simulations can reproduce the observed data. If the model produces the observed spectra, then this value should be around 50\%. For each of 10$^6$ simulations, we use parameter values drawn from best-fit model. For best fit model, 46\% realization is less than best fit statistics while it is very high $>$ 90\% if any of the model component is removed from the best-fit model. This indicates that the model which has one less component than the current best-fit model can be rejected with 90\% confidence.
In order to check whether an empirical comptonization fits better than simple powerlaw, we replace {\tt powerlaw} model with {\tt simpl} in {\tt XSpec}. The {\tt simpl} model allows to scatter a fraction of seed photons into powerlaw component. The best fit with {\tt simpl} returns reduced $\chi^2$ similar to that with simple {\tt powerlaw} model with powerlaw index of 1.70$^{+0.55}_{-0.19}$ and a scattered fraction of 0.011$^{+0.006}_{-0.009}$ when both up and down scattering are allowed. Therefore, allowing comptonization in stead of simple powerlaw does not affect our results.

The fitting of the soft state spectra allows for an estimation of the
distance to the source. The top left panel of Figure \ref{chiplots}
shows the variation of $\Delta \chi^2$ as a function of distance 
which illustrates the range of acceptable values. The normalization of the disk component depends
on the distance, the inner radius and the inclination angle. 
In order to compute $\Delta \chi^2$ variations in parameters, we use {\it steppar} in {\tt XSpec}.  

In this fitting, a distance estimate is possible, 
because the inclination angle and the inner radius are independently 
constrained being parameters of the relativistic Iron line as well 
as for the relativistic disk emission. It should be emphasized that it
is not the Iron line alone which determines these two parameters,  but
the shape of the relativistically blurred disk emission  also helps
in constraining their values. We illustrate this in the top middle panel
of Figure \ref{chiplots}. Here we fit the model by untying the inner radius for the 
disk and line components i.e., we allow them to vary independently, but
keeping the disk inclination angle of the {\tt kerrd} model tied to that of {\tt laor} model while disk inclination angle of {\tt laor} is kept free. We run {\it steppar} independently on inner disk radius parameter from {\tt kerrd} and {\tt laor} models and the result is shown in the top middle panel of \ref{chiplots}. 
The Figure shows that the variation of $\Delta \chi^2$ as a function of
the inner radius obtained from  the disk and Iron line components provide the similar estimate of the inner
radius. If instead, we fit the model by tying the inner radius of the {\tt kerrd} model component to that of the {\tt laor} model component and
let the inclination angle parameters from both model components vary independently, we get similar
constraints on disk inclination angle from both the components (by running {\tt steppar} on disk inclination angle parameters of {\tt kerrd} and {\tt laor} model components separately). The result is shown in the top right panel of Figure  
\ref{chiplots}). During the above exercises of computing $\Delta \chi^2$ variations in disk inclination angle and inner disk radius, the distance parameter was allowed to vary during the fitting. Thus, the relativistic nature of the disk emission
contributes to the estimation of the inner radius and disk inclination angle. To check whether relativistic corrections to the emission process is absolutely necessary, we use non$-$relativistic, optically thick disk models like {\tt diskbb}, {\tt diskpn} and {\tt ezdiskbb}. They either provide unusually large reduced $\chi^2$ ($>$ 2) or unacceptably high inner disk temperature ($>$ 2 keV). Therefore, relativistic treatment is required by the observed data and the requirement is consistent with both disk emission and line profile measurements.  

\subsubsection{MCMC simulations and results}

The use of $\chi^2$ minimization technique is not always reliable for the determination of model parameters with high significance when large number of free parameters are involved (for example, our best-fit spectral model has 23 free parameters including model normalization and cross calibration factors). For an independent check on our results, we employ the Markov Chain Monte Carlo (MCMC) simulations technique for validation of results obtained from $\chi^2$ minimization technique. The procedure is following : we proceed fitting the background-subtracted spectra using $\chi^2$ statistics. Since X$-$ray spectral counts usually follow Poisson distribution, we replace the fit statistic with appropriate statistic for Poisson data ({\tt pgstat} in {\tt XSpec}) which assume Poisson distribution on source spectral count but Gaussian distribution on background counts. The profile likelihood of {\tt pgstat} is derived in the same way as the likelihood of C$-$statistic is obtained. While fitting, the distance, disk inclination angle and inner disk radius parameters from {\tt kerrd} and {\tt laor} models are kept free to vary. With the new fit statistics, we run 50000 element chains starting from a random perturbation away from best-fit and ignoring first 5000 elements of the chain. The distribution of the current proposal (i.e., an assumed probability distribution for each Monte Carlo step to run the simulation) is assumed to be Gaussian with the rescaling factor of 0.001. The source of the proposal is drawn from the diagonal of the covariance matrix resulting from the fit.

\begin{figure*}
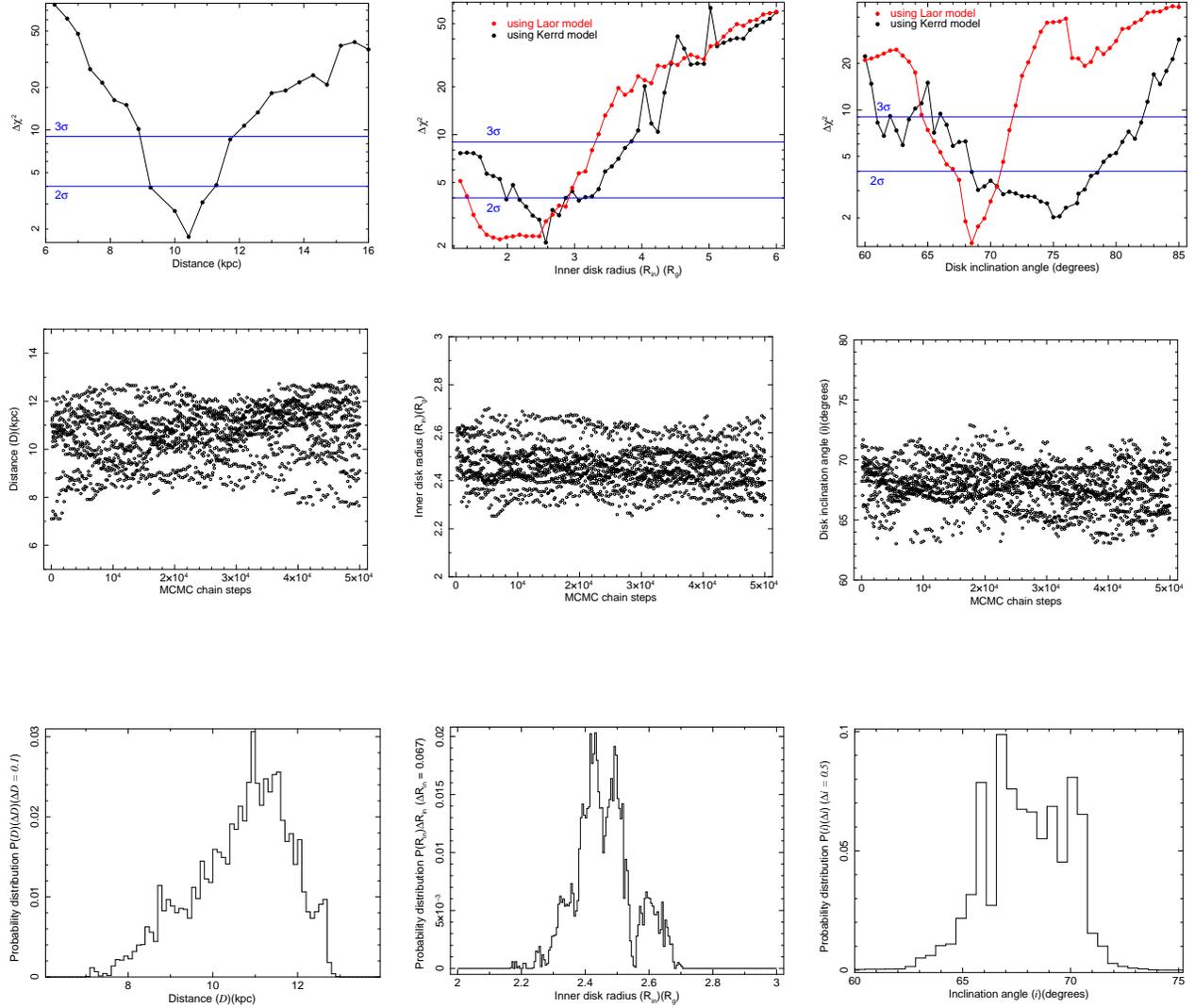

\begin{center}$
\begin{array}{ccc}
\includegraphics[width=4.5cm,height=5.5cm,angle=-90,keepaspectratio]{fig4a.ps} &
\vspace{4mm}
\includegraphics[width=4.5cm,height=5.5cm,angle=-90,keepaspectratio]{fig4b.ps} &
\vspace{1mm}
\includegraphics[width=4.5cm,height=5.5cm,angle=-90,keepaspectratio]{fig4c.ps} \\
\vspace{10mm}
\includegraphics[width=4.5cm,height=5.5cm,angle=-90,keepaspectratio]{fig4d.ps} &
\vspace{2mm}
\includegraphics[width=4.5cm,height=5.5cm,angle=-90,keepaspectratio]{fig4e.ps} &
\vspace{2mm}
\includegraphics[width=4.5cm,height=5.5cm,angle=-90,keepaspectratio]{fig4f.ps} \\
\vspace{2mm}
\includegraphics[width=4.5cm,height=5.5cm,angle=-90,keepaspectratio]{fig4g.ps} &
\vspace{2mm}
\includegraphics[width=4.5cm,height=5.5cm,angle=-90,keepaspectratio]{fig4h.ps} &
\vspace{2mm}
\includegraphics[width=4.5cm,height=5.5cm,angle=-90,keepaspectratio]{fig4i.ps}
\vspace{2mm} 
\end{array}$
\end{center}
\caption{{\bf Top panels:} Using Delta statistics, variation of $\Delta \chi^2 = \chi^2 - \chi^2_{min}$ as a function
of model parameter values are shown. The two horizontal lines mark the 2 and 3$\sigma$
confidence levels. In the top left panel, the variation in distance parameter in {\tt kerrd} model is measured while the inner
disk radius and inclination angles of {\tt kerrd} and {\tt laor} models are tied to each other. In the top middle and top right panels show the variation in inner disk radius and the disk inclination angle parameters from {\tt kerrd} and {\tt laor} models respectively when one of them is allowed to vary independently in both models keeping the other tied. {\bf Middle and bottom panels:} Results from Markov Chain Monte Carlo simulations for distance from the {\tt kerrd} model (left), inner disk radius (middle) and disk inclination angle (right) parameters from the {\tt Laor} model component of the best-fit spectral model. Parameter values along 10,000 element chain trajectory are shown in middle panels for three parameters. For clarity, every 10th element of the chain is shown. Bottom panels show 1$-$Dimensional probability distribution function for all three parameters obtained from {\tt kerrd} (distance) and {\tt Laor} (inner disk radius and disk inclination angle) model components. While running the chain for MCMC, common parameters from both {\tt kerrd} and {\tt Laor} models (i.e., inner disk radius and disk inclination angle) are allowed to vary independently.}
\label{chiplots}
\end{figure*}

Middle panels of Figure \ref{chiplots} show the parameter values of distance (from {\tt kerrd} model), inner disk radius and disk inclination angle (from {\tt laor} model) through the 50000 elements of the chain. Every 10th element of the chain is plotted for clarity. For all chain steps, highly-constrained simulated parameter values confirms that the global best-fit of the spectral parameter has been achieved. Using a different proposal (i.e., Cauchy distribution in stead of Gaussian) also converges on the same best fit with high repeatability fraction in the chain. The best-fit parameter values and MCMC-derived 2$\sigma$ errors for the soft state spectral fitting are provided in Table 1 and bottom panels of Figure \ref{chiplots} show the 1$-$dimensional probability distribution for distance (from {\tt kerrd} model), inner disk radius and disk inclination angle (from {\tt laor} model) resulting from best-fit spectral model. 

Using MCMC technique, we obtain distance to the source, D between 8.3$-$12.4 kpc, the inner disk radius, R$_{in}$ between 2.3$-$2.7 R$_g$ (from {\tt laor} model) \& $<$3.2 R$_g$ (from {\tt kerrd} model) and the disk inclination angle ({\it i}) between 64$-$72$^o$ (from {\tt laor} model) and 68-82$^o$ (from {\tt kerrd} model). Therefore, the parameters - R$_{in}$ and {\it i}, obtained from different models are consistent with moderately relativistic inner accretion flow and very large disk inclination angle. Additionally, Figure \ref{chiplots} shows that results obtained from Delta statistics (top panels) are consistent with that from MCMC simulation chains (middle and bottom panels).

The spectral analysis of the soft state using $\chi^2$ minimization as well as MCMC simulations, allows us the estimations of the distance 10.8$^{+1.6}_{-2.5}$ kpc, the inner radius of the disk $2.43^{+0.39}_{-0.17} r_g$ (average of parameters obtained from {\tt kerrd} and {\tt laor} models) and disk inclination angle of 69.5$^{+12.8}_{-4.2}$ degrees (average of parameters obtained from {\tt kerrd} and {\tt laor} models). The inner radius and the inclination angle are constrained self consistently and independently by the relativistic effects seen in both the broad Iron line and the disk emission. 

The distance estimated by the spectral fitting as well as MCMC simulation (10.8$^{+1.6}_{-2.5}$ kpc) allows us to compute the
Luminosity of  V4641 Sgr for the two observations and
subsequently the Eddington fraction. In the complete 0.3$-$30.0 keV 
energy band the luminosity during the soft state were  $4.59 \pm 0.65 \times 10^{36}$ ergs/s and $1.03 \pm 0.12 \times 10^{37}$ ergs/s considering distances to be 8.3 kpc and 12.4 kpc respectively. This corresponds
to an Eddington ratio $L/L_{Edd}$ = 12.07 $\pm$ 0.87 $\times$ 10$^{-3}$ and 5.74 $\pm$ 0.33 $\times$ 10$^{-3}$ for a $6.4 M_\odot$ black hole.

\subsection{hard state spectral analysis}

\begin{figure}
\includegraphics[scale=0.35,angle=-90,keepaspectratio]{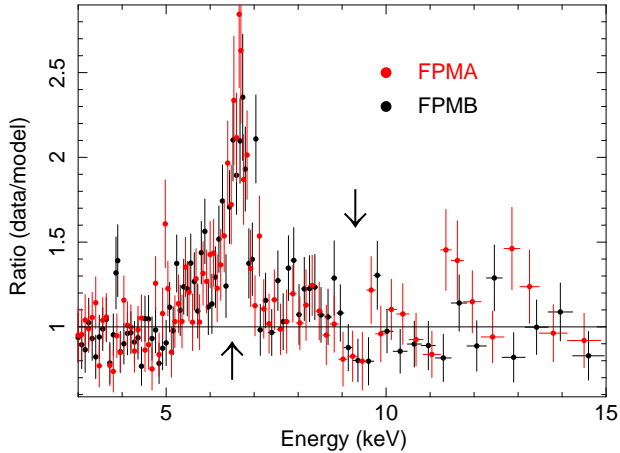} 
\caption{Data to model ratio plot of the Iron line region for the 
hard state spectrum  (on 17 April, 2014) of V4641 Sgr using
\nustar{}/FPMA (black) and \nustar{}/FPMB (red) data.
The continuum model used is a  power-law modified
by Galactic absorption. The residuals show a slightly  broad Iron line 
 and a edge at $\sim 9.5$ keV.
These features are marked by arrows. }
\label{hardresidual}
\end{figure}

For the second observation the source was harder and significantly 
less luminous. There is no evidence for a disk emission. 
An absorbed power-law fit reveals features in the 5-10 keV band as shown
in Figure \ref{hardresidual}.  There is broad line at $\sim$6 keV and
an absorption edge around $\sim$9 keV. Indeed an absorbed powerlaw,
 a slightly broad emission line at $6.5$ keV and an edge at
$8.74$ keV provides a reasonable fit with $\chi^2/dof = 284/247$.
The best fit parameters listed in Table \ref{specpar} and the 
unfolded spectrum with residuals are shown in Figure $\chi^2/dof = 284/247$ (top and middle panels). During the hard state, the luminosity was 6.91$^{+3.04}_{-2.85}$ $\times$ 10$^{35}$ ergs/s and the Eddington ratio $L/L_{Edd}$ = 7.28$^{+2.82}_{-1.71}$ $\times$ 10$^{-4}$.

\begin{figure}
\includegraphics[scale=0.35,angle=-90,keepaspectratio]{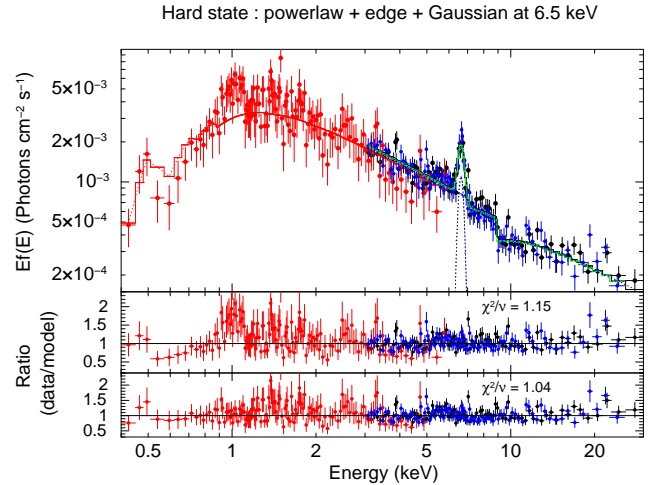} 
\caption{Top panel shows the simultaneously fitted unfolded spectra from \swift{}/XRT and
\nustar{}/FPMA \& FPMB during the hard state along with model
components and data to model ratio. The model consists of
a  powerlaw, a slightly broad Iron line and an absorption
edge. The middle panel shows the residual of the fit ($\chi^2/\nu$=284/247) where an edge like feature and a narrow line like feature are observed at $\sim$0.5 keV and $\sim$1 keV respectively. Using an additional edge as well narrow Gaussian the fit improves ($\chi^2/\nu$=252/242) as shown in the bottom panel. Similar low energy features are also seen in the soft state XRT spectrum and may be due to calibration uncertainties.}
\label{hardspec}
\end{figure}

The residual plot in the middle panel of \ref{hardspec} shows an emission line-like feature at $\sim$ 1 keV as well as edge-like feature at $\sim$0.4 keV in the XRT spectrum. To account these features, we add an edge and a Gaussian line to the best fit model. The fit improves with $\chi^2/dof = 254/242$ and the improved residual is shown at the bottom panel of Figure \ref{hardspec}. The Gaussian feature at $\sim$ 1.1 keV has the line width of $<$ 40 eV (3$\sigma$ significance) which is narrower than the instrumental response at that energy. Uncorrelated residual features are also observed at $<$ 0.5 keV from Figure \ref{residual}. These two facts indicate that the observed features are possibly due to calibration uncertainties, rather than real features from a physical process. Additionally, calibration residual at the energy $<$ 0.5 keV has been observed for \swift{}/XRT Window Timing (WT) mode data\footnote[1]{http://www.swift.ac.uk/analysis/xrt/digest$\_$cal.php$\#$abs}. Using the best fit model without the low energy edge and narrow Gaussian line, we fit the spectra excluding the energy range 0.3-1.2 keV in the \swift{}/XRT band during both soft state and hard state. In doing so, we do not observe any change in fitted parameter values within error-bars than those reported in Table 1. Reduced $\chi^2$ is also unaffected by this exercise. Therefore, these low energy features do not affect the overall fitting.

During the declining phase of 2004 outburst of V4641 Sgr, 
as the source approached quiescence, \chandra{} ACIS-S spectrum 
in 0.5-10.0 keV showed absorption-like depression between 
1-5 keV \citep{mo14} which mimics spectra from typical Seyfert-2 galaxies 
that includes partial covering and distant reflection. We fitted the
same model used by \citet{mo14} i.e. {\tt tbabs x pcfabs x
pexmon} to fit the soft state data and find unacceptable reduced
$\chi^2_{red} > 20$, even when we add an extra  Gaussian model.
For the hard state, the reduced  $\chi^2_{red}$ was found to be 1.76.
Thus our analysis shows that for both the soft and hard states
the model is not necessary.

\section{Comparison with other black hole X$-$ray binaries}

\begin{figure}
\includegraphics[scale=0.39,keepaspectratio]{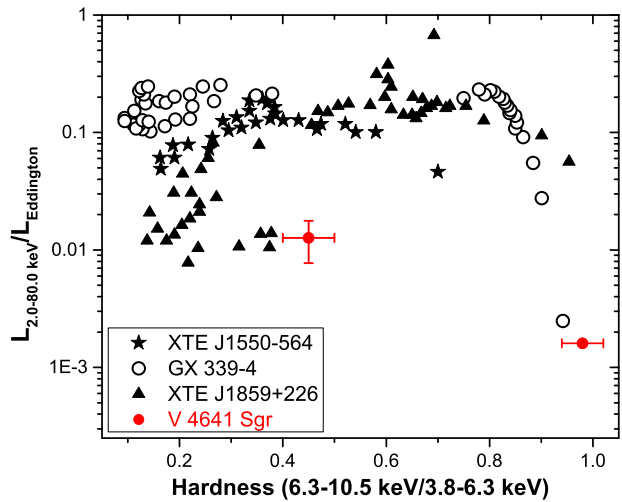}
\caption{Eddington Luminosity fractions (defined as the ratio between
the luminosity in 2.0-80.0 keV energy range and the Eddington
luminosity) as a function of Hardness (defined as the flux ratio
between 6.3$-$10.5 keV and 3.8$-$6.3 keV). The values are computed using
\xte{}/PCA data of outbursts of the low mass BHXBs: XTE J1550$-$564 (stars), GX
339$-$4 (circles), XTE J1859+226 (triangles). The plot is similar to that
obtained by \citet{fe04}. The values for the  hard and soft
state observations of V4641 Sgr using \nustar{} are
plotted using filled red circles. Note that V4641 Sgr shows a soft state
at a very low Eddington ratio $\sim 5.7 \times 10^{-3}$.}
\label{Qdiag}
\end{figure}

The fact that V4641 Sgr shows a soft state at such a low Eddington ratio
makes it unique compared to other black hole systems. To illustrate this
we compare it with three well known black hole systems : GX 339$-$4, XTE J1859+226 and XTE J1550$-$564 by reproducing the
results obtained by \citet{fe04}.  We have chosen
these systems not only because data is available for their various 
spectral states, but also their black hole masses and
distance is similar to that of V4641 Sgr \citep[see Table 1 of ][and references therein]{fe04}. For the following analysis, we considered \xte{} observations of  outbursts of these sources that occurred during 
 March 2002$-$May 2003 \citep{be04}, October$-$December 1999 \citep{br02} and August 1998$-$May
1999 \citep{so00}. Following \citet{fe04},  we fitted each of the
observations with a disk emission, power-law and a Gaussian line modified
by Galactic absorption. We used only the PCU2 of the \xte{}/PCA since it
was  the best calibrated proportional counter. We defined 
luminosity to be in 2-80 keV range and
hardness ratio (HR) as the ratio of the flux between 
 6.3$-$10.5 keV and 3.8$-$6.3 keV. The Eddington ratio versus 
hardness obtained from the analysis is shown in Figure \ref{Qdiag}. The approximate `q' shapes obtained from our analysis are consistent 
with previous analysis \citep[e.g.][and references therein]{fe04}.

In Figure \ref{Qdiag}, the points corresponding to the
hard and soft observations of V4641 Sgr are plotted for comparison.
The soft state of V4641 Sgr is significantly less luminous when compared
to the other three sources. In fact, it seems, that the position 
occupied by V4641 Sgr in the soft state, has not been observed 
in any other black hole systems. In the hard state, the Eddington
ratio is $\sim 1 \times 10^{-3}$, which is close to the ratio value below which  
black hole systems are considered to be in quiescence \citep{pl13}.

\section{Discussion}

In this paper, we have performed joint spectral analysis of
the high mass BHXB V4641 Sgr using observations from \swift{}/XRT and
\nustar{} satellites in the energy range 0.3$-$30.0 keV for two different
spectral states. While for the lower flux level hard state, the spectrum
can be simply described by a power-law and a slight broad Iron line, the
higher flux soft state reveals complex components that include a
power-law, a relativistic disk component, a relativistic broad Iron line,
 two narrow line emission and two edges.

With the 2$\sigma$ distance measurement from our analysis (10.8$^{+1.6}_{-2.5}$ kpc), one can estimate the upper and lower limit of Eddington fraction for the soft state to be $L/L_{Edd}$ = 12.07 $\pm$ 0.87 $\times$ 10$^{-3}$ and 5.74 $\pm$ 0.33 $\times$ 10$^{-3}$ respectively. The upper limit and lower limit of $L/L_{Edd}$ are estimated using upper and lower limit of distances which are 12.4 kpc and 8.3 kpc respectively. It is important to note that if the actual distance is lower than 8 kpc, then L$_{2-80}$ luminosity will diminish. As a consequence, L$_{2-80}$/L$_{Eddingon}$ will be lower. 

Such a low $L/L_{Edd}$ fraction for a soft disk dominated state is rare and matched closely with that observed from XTE J1859+226.
In the Eddington ratio hardness plot, BHXBs typically
form a `q' diagram and the soft state of  V4641 Sgr is located
at a position which can be considered as the lowest limit of three other typical BHXB systems as observed from Figure \ref{Qdiag}.
For a typical BHXB, the hard to soft transition occurs for
 $L/L_{Edd} \sim 0.50-1.0 $ and the reverse occurs when 
$L/L_{Edd} \sim 10^{-2}$.  This itself suggests that the spectral
state of a BHXB is not just a function of the accretion rate but
also depends on how the overall accretion rate is changing with time.
In one interpretation of the soft to hard
transition, a standard optically thick disk evaporates into a hot optically
thin one \citep{me94} and perhaps the rapid variability of the source does not allow for the evaporation to take place.
Such speculations require sophisticated theoretical modeling for confirmation 
and more deep observations of this source at different flux levels are required to understand its temporal behavior.

The inner disk radius during the soft state is found to be
$r_{in}  \sim 2.5 r_g$. If identified with the last stable orbit
of a disk, this would suggest that the black hole has a moderate
spin. However, we note that both the {\it laor} and {\it kerrd} spectral
models used for the disk and line emissions, assume an extremal
spin of $0.998$ and hence the actual value of the inner radius
maybe different if the spin is moderate. A more general prescription
which takes into different spin values of the black hole maybe used to
fit the disk emission  as has been done for some black hole systems \citep{mc14}. In such a analysis, one should also
consider more realistic color factors than the universal $1.7$ assumed
in this work. However such analysis has been undertaken on good quality clean data where the power-law component
is very weak i.e. $<$10\%. Nevertheless, it will be interesting
to see whether the spin of the black hole can be constrained using
sophisticated spectral models, an analysis which is beyond the scope
of the present work.

\subsection{Effects of companion star on observed spectra}

it is important to explore the role of strong wind from the companion star in explaining observed features in the spectra. Optical depth of Ni edge is during the soft state observation is found to be unusually large and vary significantly while transiting from the soft state to the hard state (e.g., for the absorption edge at $\sim$9 keV, $\tau$ increases from $\sim$ 0.22 to $\sim$ 0.47). Secondly, during the soft state, two narrow emission lines and edges are clearly detected in the spectrum. Since the lines are narrow, they should arise from regions far from the black hole. Therefore, the presence of highly ionized wind from the disk/binary companion seems to be a natural origin. The emission line at $\sim 6.95$ keV  can be identified due to ionized Iron in the X$-$ray irradiated wind. The emission line at $\sim 8.1$ keV is more uncertain. The line energy suggests that it could arise due to highly ionized Nickel as can be seen in Figure 2 of \citet{tu92}.

However, the nature of wind need to be investigated carefully. A relationship between the kinetic power of the wind (per unit filling factor, C$_v$) and the bolometric luminosity, log(L$_{wind,42}$/C$_v$) = 1.58$\pm$0.07 log(L$_{bol,42}$) - 3.19$\pm$0.19, is obtained by \citet{ki13} by studying the \chandra{} grating spectra across black hole mass scale $-$ from BHXBs to AGNs. If the above relationship is true for V4641 Sgr, than using the bolometric luminosity for the soft spectral state in our analysis, we obtain log(L$_{wind,42}$/C$_v$) = -2.46 which significantly deviates from the value observed for canonical BHXBs and consistent with low ionization. In order to produce large optical depth of edges, circumstellar wind from the binary companion in V4641 Sgr need to be very strong. From the observation, the mass of the binary companion of V4641 Sgr is found out to be 2.9 $\pm$ 0.4 M$\odot$ which is a reddened B9III star with relatively low rotational velocity (V$_{rot}$sin{\it i} = 100.9 $\pm$ 0.08 km/s). The measured velocity is lower than the predicted minimum threshold for launching highly ionized stellar wind (V$_{rot}$sin{\it i} of 150$-$200 km/s from the UV survey of 40 B-type stars covering luminosity classes V-III  \citep{gr89}). Secondly from the spectroscopic study, the possibility of radiation driven wind from the surface of the companion star can be ruled out since the surface temperature of the companion star in V4641 Sgr is derived to be 10250 $\pm$ 300 K \citep{ma14} which is relatively low and not hot enough to produce an effective radiation driven wind. This indicates that the possibility of very strong wind in this source may not be the only explanation of observed features. It is suggestive that multiple and strong emission lines and absorption edges, even at hard state may possibly indicate the over-abundance of elements like Fe (A$_{Fe}$ $>$ 2) present in the spectra. However, high resolution grating spectroscopy using \chandra{} or \xmm{} during soft state in V4641 Sgr may provide compelling evidence for any possible nature of disk wind in this source. 

It is interesting to note that
for the Galactic jet source SS 433, the Ni over-abundance by a factor of 20 \citep{ko97} and high disk inclination angle causes the visibility of Ni emission line in the spectra \citep{ko97} and Nickel
emission lines are expected from ionized medium surrounding Active
Galactic Nuclei \citep{tu92}. A detailed work focusing on the
physics of the X$-$ray irradiated wind is required to understand
the narrow emission lines and edges seen in V4641 Sgr. Although, there
does not seem to be any issues regarding \nustar{} calibration during
these observations, it would be perhaps be prudent to reconfirm these
emission lines with a second observation by \nustar{} during similar spectral state or by other
high resolution instruments.

\section{Acknowledgments}

We thank the referee for useful discussions and comments. We are thankful to \nustar{} team for making data publicly available. This research has made use of the \nustar{} Data Analysis Software (NuSTARDAS) jointly developed by the ASI Science Data Center (ASDC, Italy) and the California Institute of Technology (USA). \xte{}/PCA data obtained through the High Energy Astrophysics Science Archive Research Center online service, provided by the NASA/Goddard Space Flight Center and the SWIFT data center. \maxi{} data are obtained from MAXI team, RIKEN, JAXA.

\clearpage

\begin{table*}
 \centering
 \caption{Observation and spectral fitting details of V4641 Sgr using simultaneous data from \swift{}/XRT and \nustar{} during soft
 state and hard state. The best fit model of the soft state spectrum consists of optically thick, relativistic accretion disk model ({\tt kerrd}), relativistic line emission model ({\tt laor}), powerlaw component ({\tt powerlaw}), two narrow emission lines modeled by two Gaussians ({\tt ga}) and two absorption features modeled by two edges ({\tt edge}). {\tt TBabs} model is used for Galactic absorption. For hard state, {\tt powerlaw} model is sufficient to fit the continuum spectrum. Cross calibration between different instruments are taken care by {\tt const}. In the table, N$_H$ is the neutral Hydrogen column density, $\Gamma_p$ is the powerlaw index, E$_{edge1}$ and E$_{edge2}$ are two absorption edge energies and $\tau_{edge1}$ and $\tau_{edge2}$ are their optical depths respectively. E$_{L}$ is the Laor line energy and W$_{eq,L}$ is its equivalent width, q$_{L}$ is the emissivity index. R$_{in,L}$ and R$_{out,L}$ are inner and outer disk radius from {\tt laor} model. R$_{in,kerrd}$ is the inner disk radius from {\tt kerrd} model. {\it i$_L$} and {\it i$_{kerrd}$} are the disk inclination angle obtained from {\tt laor} and {\tt kerrd} models respectively. While fitting, no two similar parameters are tied. E$_{Fe}$ is the narrow iron line energy, W$_{eq,Fe}$ is the equivalent width of the narrow Fe line, E$_{Ni}$ is the Ni K$_\alpha$ line energy and W$_{eq,Ni}$ is its equivalent width. $\dot{M}$$_{acc}$ is the mass accretion rate, D is the distance to the source, F$_{total}$, F$_{laor}$ and F$_{power}$ are total flux, fluxes due to {\tt laor} and {\tt power} model components. MCMC derived 2$\sigma$ errors are quoted for parameters.}
\begin{center}
\scalebox{0.97}{%
\begin{tabular}{ccc}
\hline 
Parameter & soft state  & hard state \\
\hline
Date & 2014-02-14 & 2014-04-17 \\
\hline
Obs ID & & \\
\hline 
Nustar & 80002012002 & 80002012004 \\
XRT  & 00030111030 & 00030111046 \\
\hline
Exposure time (sec) & & \\
\hline
Nustar & 26420 & 24045 \\
(DT corrected) & & \\
XRT & 1944 & 3983 \\
\hline
Model & tbabs*edge*edge*(powerlaw+ & tbabs*edge*(powerlaw+ga) \\
      &  kerrd+laor+ga+ga) &  \\
\hline
N$_H$ & & \\
(10$^{22}$ cm$^{-2}$) & 0.21$^{+0.02}_{-0.04}$ & 0.31$^{+0.04}_{-0.06}$ \\ [0.2cm]
$\Gamma_{p}$ & 1.79$^{+0.59}_{-0.22}$ & 2.12$^{+0.05}_{-0.06}$ \\[0.2cm]
E$_{edge1}$ (keV) & 7.21$^{+0.05}_{-0.14}$ & -- \\ [0.2cm]
$\tau_{edge1}$ & 0.21$^{+0.03}_{-0.11}$ & -- \\ [0.2cm]
E$_{edge2}$ (keV) & 9.61$^{+0.03}_{-0.11}$ & 8.74$^{+0.21}_{-0.18}$ \\ [0.2cm]
$\tau_{edge2}$ & 0.22$^{+0.07}_{-0.04}$ & 0.47$^{+0.11}_{-0.09}$ \\ [0.2cm]
E$_{L}$ (keV) & 6.36$^{+0.28}_{-0.08}$ & -- \\[0.2cm]
q$_{L}$ & 2.02$^{+0.55}_{-0.05}$ & -- \\[0.2cm]
R$_{in,L}$ (R$_g$) & 2.43$^{+0.23}_{-0.17}$ & -- \\[0.2cm]
R$_{in,kerrd}$ (R$_g$) & 2.56$^{+0.55}_{-0.54}$ & -- \\[0.2cm]
R$_{out,L}$ (R$_g$) & 400 (fixed) & -- \\[0.2cm]
{\it i$_L$} (degrees) & 69.5$^{+4.2}_{-4.6}$ & -- \\[0.2cm]
{\it i$_{kerrd}$} (degrees) & 76.1$^{+5.2}_{-10.3}$ & -- \\[0.2cm]
W$_{eq,L}$ (eV) & 292.8$^{+66.6}_{-46.5}$ & -- \\[0.2cm]
E$_{Fe}$ (keV) & 6.95$^{+0.03}_{-0.02}$ & 6.61$^{+0.04}_{-0.07}$ \\[0.2cm]
W$_{eq,Fe}$ (eV) & 140.4$^{+18.5}_{-22.4}$ & 526.8$^{+54.7}_{-72.5}$ \\[0.2cm]
E$_{Ni}$ (keV) & 8.31$^{+0.05}_{-0.14}$ & -- \\[0.2cm]
W$_{eq,Ni}$ (eV) & 193.6$^{+24.7}_{-56.7}$ & -- \\[0.2cm]
$\dot{M}$$_{acc}$ & & \\
(10$^{18}$ gm/s) & 0.065$^{+0.015}_{-0.018}$ & -- \\ [0.2cm]
D (kpc) & 10.8$^{+1.6}_{-2.5}$ & -- \\ [0.2cm]
F$_{total}$ (10$^{-10}$ ergs/s/cm$^{2}$) & 5.87$^{+0.11}_{-0.08}$ & 0.36$^{+0.04}_{-0.06}$ \\[0.2cm]
F$_{laor}$ (10$^{-11}$ ergs/s/cm$^{2}$) & 1.25$^{+0.06}_{-0.07}$ & -- \\[0.2cm]
F$_{power}$ (10$^{-11}$ ergs/s/cm$^{2}$)  & 2.41$^{+0.11}_{-0.04}$ & -- \\[0.2cm]
$\chi^2$/$\nu$ & 721/696 & 284/247 \\
\hline
\end{tabular}}
\end{center}
\label{specpar}
\end{table*}

\clearpage

\end{document}